\RequirePackage{fix-cm}
\documentclass[twocolumn,epjc3]{svjour3}
\smartqed  
\usepackage{amsmath}
\usepackage{amssymb}
\usepackage{amssymb,bbold}
\usepackage{kantlipsum,widetext}

\RequirePackage{graphicx}
\RequirePackage[colorlinks,citecolor=blue,urlcolor=blue,linkcolor=blue]{hyperref}
\RequirePackage[numbers,sort&compress]{natbib}

\journalname{Eur. Phys. J. C}
\begin{document}

\title{Quantum dynamics of scalar bosons in a cosmic string background
}


\author{Luis B. Castro\thanksref{e1,addr1} 
}

\thankstext{e1}{e-mail: luis.castro@ufma.br}


\institute{Departamento de F\'{\i}sica, Universidade Federal do Maranh\~{a}o, Campus Universit\'{a}rio do Bacanga, 65080-805, S\~{a}o Lu\'{\i}s, MA, Brazil.
\label{addr1}
}

\date{Received: date / Accepted: date}

\maketitle

\begin{abstract}
The quantum dynamics of scalar bosons embedded in the background of a cosmic string is considered. In this work, scalar bosons are described by the Duffin-Kemmer-Petiau (DKP) formalism. In particular, the effects of this topological defect in the equation of motion, energy spectrum and DKP spinor are analyzed and discussed in details. The exact solutions for the DKP oscillator in this background are presented in a closed form.

\end{abstract}

\section{Introduction}
\label{intro}

The first-order Duffin-Kemmer-Petiau (DKP) formalism \cite{Petiau1936,Kemmer1938,PR54:1114:1938,Kemmer1939}
describes spin-zero and spin-one particles and has been used to analyze relativistic interactions of spin-zero and spin-one hadrons with nuclei as an alternative to their conventional second-order Klein-Gordon (KG) and Proca counterparts. Although the formalisms are equivalent in the case of minimally coupled vector interactions \cite{PLA244:329:1998,PLA268:165:2000,PRA90:022101:2014}, the DKP formalism enjoys a richness of couplings not capable of being expressed in the KG and Proca theories \cite{PRD15:1518:1977,JPA12:665:1979}. Recently, there has been an increasing interest on the so-called DKP oscillator \cite{ZPC56:421:1992,JPA27:4301:1994,JPA31:3867:1998,JPA31:6717:1998,PLA346:261:2005,
JMP47:062301:2006,PS76:669:2007,PS78:045010:2008,JMP49:022302:2008,
IJTP47:2249:2008}. The DKP oscillator considering minimal length \cite{JMP50:023508:2009,
JMP51:033516:2010} and noncommutative phase space \cite{CTP50:587:2008,CJP87:989:2009,IJTP49:644:2010,EPJC72:2217:2012} have also appeared in the literature. The DKP oscillator is a kind of tensor coupling with a linear potential which leads to the harmonic oscillator problem in the weak-coupling limit. Also, a sort of vector DKP oscillator (non-minimal vector coupling with a linear potential \cite{MPLA20:43:2005,JPA43:055306:2010,NPBPS199:203:2010,PLA375:2596:2011,
ADHEP2014:784072:2014} has been an topic of recent investigation. Vector DKP oscillator is the name given to the system with a Lorentz vector coupling which exhibits an equally spaced energy spectrum in the weak-coupling limit.
The name distinguishes from that system called DKP oscillator with Lorentz tensor couplings of Ref.~\cite{ZPC56:421:1992,JPA27:4301:1994,JPA31:3867:1998,JPA31:6717:1998,PLA346:261:2005,
JMP47:062301:2006,PS76:669:2007,PS78:045010:2008,JMP49:022302:2008,
IJTP47:2249:2008,JMP50:023508:2009,JMP51:033516:2010,CTP50:587:2008,CJP87:989:2009,IJTP49:644:2010,
EPJC72:2217:2012}.

The DKP oscillator is an analogous to Dirac oscillator \cite{JPA22:L817:1989}. The Dirac oscillator is a natural model for studying properties of physical systems, it is an exactly solvable model, several research have been developed in the context of this theoretical framework in recent years. A detailed description for the Dirac oscillator is given in Ref.~\cite{STRANGE1998} and for other contributions see Refs.~\cite{PLA325:21:2004,JPA39:5125:2006,PRC73:054309:2006,
PRA84:032109:2011,AP336:489:2013,PLB731:327:2014,AP356:83:2015}. The Dirac oscillator embedded in a cosmic string background has inspired a great deal of research in last years\cite{NPB328:140:1989,PRL62:1071:1989,PLA361:13:2007,PRD83:125025:2011,
PRD85:041701:2012,AP339:510:2013,EPJC74:3187:2014}. A cosmic string is a linear defect that change the topology of the medium when viewed globally. The influence of this topological defect in the dynamics of spin-$1/2$ particles has been widely discussed in the literature. However, the same problem involves bosons via DKP formalism has not been established. Therefore, we believe that this problem deserves to be explored.

The main motivation of this work is inspired by the results obtained in Ref.~\cite{EPJC74:3187:2014}. As a natural extension, we address the quantum dynamics of scalar bosons (via DKP formalism) embedded in the background of a cosmic string. The influence of this topological defect in the equation of motion, energy spectrum and DKP spinor are analyzed and discussed in details. The case of DKP oscillator in this background is also considered. In this case, the problem is mapped into a Schr\"{o}dinger-like equation embedded in a three-dimensional harmonic oscillator for the first component of the DKP spinor and the remaining components are expressed in terms of the first one in a simple way. Our results are very similar to Dirac oscillator in a cosmic string background, except by the absence of terms that depend on the spin projection parameter.

This work is organized as follows. In section \ref{sec:1}, we consider the DKP equation in a curved space-time. We discuss conditions on the interactions which lead to a conserved current in a curved space-time (section \ref{subsec:1}). In section \ref{sec:2}, we give a brief review on a cosmic string background and we also analyze the curved-space beta matrices and spin connection in this background. In section \ref{sec:3}, we concentrate our efforts in the interaction called DKP oscillator embedded in the background of a cosmic string. In particular, we focus the case of scalar bosons and obtain the equation of motion, energy spectrum and DKP spinor (section \ref{subsec:3}). Finally, in section \ref{sec:4} we present our conclusions.

\section{Duffin-Kemmer-Petiau equation in a curved space-time}
\label{sec:1}

The Duffin-Kemmer-Petiau (DKP) equation for a free boson in curved space-time is given by \cite{GRG34:491:2002,
GRG34:1941:2002} ($\hbar =c=1$)%
\begin{equation}\label{dkp}
\left[ i\beta ^{\mu }\nabla_{\mu}-M\right] \Psi =0
\end{equation}%
\noindent where the covariant derivative
\begin{equation}\label{der_cov}
\nabla_{\mu}=\partial _{\mu }-\Gamma_{\mu}\,.
\end{equation}
\noindent The affine connection is defined by
\begin{equation}\label{affine}
\Gamma_{\mu}=\frac{1}{2}\,\omega_{\mu\bar{a}\bar{b}}[\beta^{\bar{a}},\beta^{\bar{b}}]\,.
\end{equation}
\noindent The curved-space beta matrices are
\begin{equation}\label{beta_curved}
\beta ^{\mu }=e^{\mu}\,_{\bar{a}}\,\beta^{\bar{a}}
\end{equation}
\noindent and satisfy the algebra%
\begin{equation}\label{betaalge}
\beta ^{\mu }\beta ^{\nu }\beta ^{\lambda }+\beta ^{\lambda }\beta
^{\nu }\beta ^{\mu }=g^{\mu \nu }\beta ^{\lambda }+g^{\lambda \nu
}\beta ^{\mu }\,.
\end{equation}%
\noindent where $g^{\mu \nu }$ is the metric tensor. The algebra expressed by (\ref{betaalge}) generates a set of 126 independent matrices whose irreducible representations are a trivial representation, a five-dimensional representation describing the spin-zero particles (scalar sector) and a ten-dimensional representation associated to spin-one particles (vector sector). The DKP spinor has an excess of components and the theory has to be supplemented by an equation which allows one to eliminate the superfluous components. That constraint equation is obtained by multiplying the Eq. (\ref{dkp}) by $1-\beta^{0}\beta^{0}$ from the left, namely
\begin{equation}\label{vinculo}
i\beta^{j}\beta^{0}\beta^{0}\nabla_{j}\Psi=M\left(1-\beta^{0}\beta^{0}\right)\Psi
\end{equation}
\noindent This constraint equation expresses three (four) components of the spinor by the other two (six) components and their space covariant derivate in the scalar (vector) sector so that the redundant components disappear and there only remain the physical components of the DKP theory.

The \textit{tetrads} $e_{\mu}\,^{\bar{a}}(x)$ satisfy the relations%
\begin{equation}\label{tetr1}
\eta^{\bar{a}\bar{b}}=e_{\mu}\,^{\bar{a}}\,e_{\nu}\,^{\bar{b}}\,g^{\mu\nu}
\end{equation}%
\begin{equation}\label{tetr2}
g_{\mu\nu}=e_{\mu}\,^{\bar{a}}\,e_{\nu}\,^{\bar{b}}\,\eta_{\bar{a}\bar{b}}
\end{equation}%
\noindent and
\begin{equation}\label{tetr3}
e_{\mu}\,^{\bar{a}}\,e^{\mu}\,_{\bar{b}}=\delta^{\bar{a}}_{\bar{b}}
\end{equation}%
\noindent the Latin indexes being raised and lowered by the Min\-kowski metric tensor $\eta^{\bar{a}\bar{b}}$ with signature $(-,+,+,+)$ and the Greek ones by the metric tensor $g^{\mu\nu}$.

\noindent The spin connection $\omega_{\mu\bar{a}\bar{b}}$ is given by
\begin{equation}\label{con}
\omega_{\mu}\,^{\bar{a}\bar{b}}=e_{\alpha}\,^{\bar{a}}\,e^{\nu \bar{b}}\,\Gamma_{\mu\nu}^{\alpha}
-e^{\nu \bar{b}}\partial_{\mu}e_{\nu}\,^{\bar{a}}
\end{equation}%
\noindent with $\omega_{\mu}\,^{\bar{a}\bar{b}}=-\omega_{\mu}\,^{\bar{b}\bar{a}}$ and $\Gamma_{\mu\nu}^{\alpha}$ are the Christoffel symbols given by
\begin{equation}\label{symc}
\Gamma_{\mu\nu}^{\alpha}=\frac{g^{\alpha\beta}}{2}\left( \partial_{\mu}g_{\beta\nu}+
\partial_{\nu}g_{\beta\mu}-\partial_{\beta}g_{\mu\nu} \right).
\end{equation}%
\noindent In this stage, it is useful to consider the current. The conservation law for $J^{\mu}$ follows from the standard procedure of multiplying (\ref{dkp}) and its complex conjugate by $\bar{\Psi}$ from the left and by $\eta^{0}\Psi$ from the right, respectively. The sum of those resulting equations leads to
\begin{equation}\label{corr}
\nabla_{\mu}J^{\mu}=\frac{1}{2}\bar{\Psi}\left(\nabla_{\mu}\beta^{\mu}\right)\Psi
\end{equation}
\noindent where $J^{\mu}=\frac{1}{2}\bar{\Psi}\beta^{\mu}\Psi$. The factor $1/2$ multiplying $\bar{\Psi}\beta^{\mu}\Psi$, of no importance regarding the conservation law, is in order to hand over a charge density conformable to that one used in the KG theory and its
nonrelativistic limit \cite{JPA43:055306:2010}. The adjoint spinor $\bar{\Psi}$ is given by $\bar{\Psi}=\Psi^{\dagger }\eta ^{0}$ with $\eta ^{0}=2\beta ^{0}\beta ^{0}-1$ in such a way
that $(\eta^{0}\beta^{\mu})^{\dag}=\eta^{0}\beta^{\mu}$ (the matrices $\beta^{\mu}$ are Hermitian with respect to $\eta^{0}$). Despite the
similarity to the Dirac equation, the DKP equation involves singular
matrices, the time component of $J^{\mu}$ is not positive definite and the
case of massless bosons cannot be obtained by a limiting process~\cite{PRD10:4049:1974}%
. Nevertheless, the matrices $\beta^{\mu}$ plus the unit operator generate a
ring consistent with integer-spin algebra and $J^{0}$ may be interpreted as
a charge density. Thus, if
\begin{equation}\label{cj0}
    \nabla_{\mu}\beta^{\mu}=0
\end{equation}
\noindent then four-current will be conserved. The condition (\ref{cj0}) is the purely geometrical assertion that the curved-space beta matrices are covariantly constant.

The normalization condition $\int{d\tau J^{0}}=\pm 1$ can be expressed as
\begin{equation}\label{normali1}
\int{d\tau \bar{\Psi}\beta^{0}\Psi}=\pm 2\,,
\end{equation}
\noindent where the plus (minus) sign must be used for a positive (negative) charge.

\subsection{Interaction in the Duffin-Kemmer-Petiau equation}
\label{subsec:1}

With the introduction of interactions, the DKP equation in a curved space-time can be written as%
\begin{equation}
\left( i\beta ^{\mu }\nabla _{\mu }-m-U\right) \Psi =0  \label{dkp2}
\end{equation}
\noindent where the more general potential matrix $U$ is written in terms of
25 (100) linearly independent matrices pertinent to five (ten)-dimensional
irreducible representation associated to the scalar (vector) sector. In the
presence of interaction, $J^{\mu }$ satisfies the equation
\begin{equation}
\nabla _{\mu }J^{\mu }+\frac{i}{2}\,\bar{\Psi}\left( U-\eta ^{0}U^{\dagger
}\eta ^{0}\right) \Psi =\frac{1}{2}\bar{\Psi}\left(\nabla_{\mu}\beta^{\mu}\right)\Psi  \label{corrent2}
\end{equation}
\noindent Thus, if $U$ is Hermitian with respect to $\eta ^{0}$ and the curved-space beta matrices are covariantly constant then four-current will be conserved. The potential matrix $U$ can be written in terms of well-defined Lorentz structures. For the spin-zero sector there are
two scalar, two vector and two tensor terms \cite{PRD15:1518:1977}, whereas for the
spin-one sector there are two scalar, two vector, a pseudoscalar, two
pseudovector and eight tensor terms \cite{JPA12:665:1979}. The condition (\ref{corrent2}) for the case of Minkowski space-time has been used to point out a
misleading treatment in the recent literature regarding analytical solutions
for nonminimal vector interactions \cite{ADHEP2014:784072:2014,CJP87:857:2009,CJP87:1185:2009,JPA45:075302:2012}.

\section{Cosmic string background}
\label{sec:2}

The cosmic string space-time is an object described by the line element
\begin{equation}\label{metric}
ds^{2}=-dt^{2}+dr^{2}+\alpha^{2}r^{2}d\varphi^{2}+dz^{2}
\end{equation}
\noindent in cylindrical coordinates $(t,r,\varphi,z)$, where $-\infty<z<+\infty$, $r\geq 0$ and $0\leq\varphi\leq2\pi$. The parameter $\alpha$ is associated with the linear mass density $\tilde{m}$ of the string by $\alpha=1-4\tilde{m}$ and runs in the interval $\left( 0,1 \right]$ and corresponds to a deficit angle $\gamma=2\pi(1-\alpha)$. In the geometric context, the line element (\ref{metric}) is related to a Minkowski space-time with a conical singularity \cite{JHEP2004:016:2004}. Note that, in the limit as $\alpha\rightarrow1$ we obtain the line element of cylindrical coordinates.

The basis tetrad $e^{\mu}\,_{\bar{a}}$ from the line element (\ref{metric}) is chosen to be
\begin{equation}\label{tetra}
e^{\mu}\,_{\bar{a}}=%
\begin{pmatrix}
1 & 0 & 0 &  0\\
0 & \cos{\varphi} & \sin{\varphi} & 0\\
0 & -\frac{\sin{\varphi}}{\alpha r} & \frac{\cos{\varphi}}{\alpha r} & 0\\
0 & 0 & 0 & 1
\end{pmatrix}%
\,.
\end{equation}%
\noindent For the specific basis tetrad (\ref{tetra}) the curved-space beta matrices read
\begin{eqnarray}
\beta^{0} &=& \beta^{\bar{0}}\,,  \label{betat} \\
\beta^{r} &=& \beta^{\bar{1}}\cos{\varphi}+\beta^{\bar{2}}\sin{\varphi}\,,   \label{betar} \\
\beta^{\varphi} &=& \frac{-\beta^{\bar{1}}\sin{\varphi}+\beta^{\bar{2}}\cos{\varphi}}{\alpha r}\,,  \label{betaphi}\\
\beta^{z}&=& \beta^{\bar{3}}\,, \label{betaz}
\end{eqnarray}%
\noindent and the spin connection is given by
\begin{equation}\label{gphi}
\Gamma_{\varphi}=(1-\alpha)\left[\beta^{\bar{1}},\beta^{\bar{2}}\right]\,.
\end{equation}
\noindent Thereby, the covariant derivative gets
\begin{eqnarray}
\nabla_{0} &=& \partial_{0}\,,\label{dt}\\
\nabla_{r} &=& \partial_{r}\,,\label{dr}\\
\nabla_{\varphi} &=& \partial_{\varphi}-(1-\alpha)\left[\beta^{\bar{1}},\beta^{\bar{2}}\right]\,,\label{dphi}\\
\nabla_{z} &=& \partial_{z}
\end{eqnarray}

\noindent Now we focus attention on the condition (\ref{cj0}) for a cosmic string background. Using the line element (\ref{metric}) and the representation for the curved-space beta matrices (\ref{betat}), (\ref{betar}), (\ref{betaphi}) and (\ref{betaz}) the condition (\ref{cj0}) is satisfied and therefore the current is conserved for this background.
Having set up the DKP equation in a cosmic string background, we are now in a position to use the machinery developed above in order to solve the DKP equation in this background with some specific forms for external interactions.

\section{DKP oscillator in a cosmic string background}
\label{sec:3}

In this section, we concentrate our efforts in the interaction called DKP oscillator embedded in the background of a cosmic string. For this external interaction we use the non-minimal substitution \cite{JPA27:4301:1994}
\begin{equation}
\vec{p}\rightarrow \vec{p}-iM\omega\eta^0\vec{r}
\end{equation}
\noindent where $\omega$ is the oscillator frequency. This interaction is a Lorentz-tensor type and is Hermitian with respect to $\eta^0$, so it furnishes a conserved four-current. Considering only the radial component the non-minimal substitution gets
\begin{equation}
\vec{p}\rightarrow \vec{p}-iM\omega\eta^0r\hat{r}\,.
\end{equation}
\noindent As the interaction is time-independent one can write $\Psi(\vec{r},t)=\Phi(\vec{r})\mathrm{exp}\left(-iEt\right)$, where $E$ is the energy of the scalar boson, in such a way that the time-independent DKP equation becomes
\begin{equation}
\left[ \beta^{\bar{0}}E+i\beta^{r}\left( \partial_{r}+M\omega\eta^{0}r \right)+
i\beta^{\varphi}\nabla_{\varphi}+
i\beta^{\bar{3}}\partial_{z}-M \right]\Phi=0
\end{equation}
\noindent where $\beta^{r}$, $\beta^{\varphi}$ and $\nabla_{\varphi}$ are given by (\ref{betar}), (\ref{betaphi}) and (\ref{dphi}), respectively.

\subsection{Scalar sector}
\label{subsec:3}

For the case of scalar bosons (scalar sector), we use the standard representation for the beta matrices given by \cite{JPG19:87:1993}
\begin{equation}
\beta ^{\bar{0}}=%
\begin{pmatrix}
\theta & \overline{0} \\
\overline{0}^{T} & \mathbf{0}%
\end{pmatrix}%
,\quad \overrightarrow{\beta }=%
\begin{pmatrix}
\widetilde{0} & \overrightarrow{\rho } \\
-\overrightarrow{\rho }^{\,T} & \mathbf{0}%
\end{pmatrix}%
\end{equation}%
\noindent where%
\begin{eqnarray}
\ \theta &=&%
\begin{pmatrix}
0 & 1 \\
1 & 0%
\end{pmatrix}%
,\quad \rho ^{1}=%
\begin{pmatrix}
-1 & 0 & 0 \\
0 & 0 & 0%
\end{pmatrix}
\notag \\
&& \\
\rho ^{2} &=&%
\begin{pmatrix}
0 & -1 & 0 \\
0 & 0 & 0%
\end{pmatrix}%
,\quad \rho ^{3}=%
\begin{pmatrix}
0 & 0 & -1 \\
0 & 0 & 0%
\end{pmatrix}
\notag
\end{eqnarray}
\noindent $\overline{0}$, $\widetilde{0}$ and $\mathbf{0}$ are 2$\times $3, 2%
$\times $2 and 3$\times $3 zero matrices, respectively, while the
superscript T designates matrix transposition. The five-component spinor can be written as $\Phi ^{T}=\left( \Phi_{1},...,\Phi _{5}\right)$ and the DKP equation for scalar bosons becomes
\begin{eqnarray}
E\Phi_{2}&-&M\Phi_{1}-i\left(\partial_{-}-\delta_{\alpha}\cos\varphi\right)\Phi_{3}\notag\label{phi1}\\
&-&i\left(\partial_{+}-\delta_{\alpha}\sin\varphi\right)\Phi_{4}-i\partial_{z}\Phi_{5}=0\,,\\
\Phi_{2} &=& \frac{E}{M}\Phi_{1} \,,  \label{phi2} \\
\Phi_{3} &=& \frac{i}{M}\left(\partial_{-}+M\omega r\cos{\varphi}\right)\Phi_{1} \,,   \label{phi3} \\
\Phi_{4} &=& \frac{i}{M}\left(\partial_{+}+M\omega r\sin{\varphi}\right)\Phi_{1}  \,,  \label{phi4}\\
\Phi_{5}&=& \frac{i}{M}\partial_{z}\Phi_{1} \,, \label{phi5}
\end{eqnarray}%
\noindent where
\begin{eqnarray}
\partial_{-}&=&\cos{\varphi}\partial_{r}-\frac{\sin{\varphi}}{\alpha r}\partial_{\varphi}\,,\label{dplus}\\
\partial_{+}&=&\sin{\varphi}\partial_{r}+\frac{\cos{\varphi}}{\alpha r}\partial_{\varphi} \label{dminus}
\end{eqnarray}
\noindent and
\begin{equation}\label{delta}
\delta_{\alpha}=\frac{1-\alpha}{\alpha r}+M\omega r\,.
\end{equation}
\noindent Meanwhile,
\begin{equation}\label{norma2}
J^{0}=\mathrm{Re}\left(\Phi_{2}^{\ast}\Phi_{1}\right)=\frac{E}{M}|\Phi_{1}|^{2}\,.
\end{equation}

\noindent Combining these results we obtain a equation of motion for the first component of the DKP spinor
\begin{equation}\label{ed2o1}
\left[\nabla_{\alpha}^2-M^2\omega^2r^2+E^2-M^2+2M\omega\right]\Phi_{1}=0
\end{equation}
\noindent where $\nabla_{\alpha}^2$ is the Laplacian-Beltrami operator in the conical space and is given by
\begin{equation}\label{laplacc}
\nabla_{\alpha}^2=\frac{1}{r}\frac{\partial}{\partial r}\left(r\frac{\partial}{\partial r}\right)+
\frac{1}{\alpha^2r^2}\frac{\partial^2}{\partial \varphi^2}+\frac{\partial^2}{\partial z^2}\,.
\end{equation}
\noindent At this stage, we can use the invariance under boosts along the $z$-direction and adopt the usual decomposition
\begin{equation}\label{ansatz}
\Phi_{1}(r,\varphi,z)=\frac{\phi_{1}(r)}{\sqrt{r}}e^{im\varphi+ik_{z}z}
\end{equation}
\noindent with $m\in\mathbb{Z}$. Inserting this into Eq. (\ref{ed2o1}), we get
\begin{equation}\label{ed2o}
\left[\frac{d^2}{dr^2}-\lambda^2r^2-\frac{\left(m_{\alpha}^2-\frac{1}{4}\right)}{r^2}+\kappa^2\right]\phi_{1}=0
\end{equation}
\noindent where $m_{\alpha}=m/\alpha$, $\lambda=M\omega$ and
\begin{equation}
\kappa=\sqrt{E^2-M^2+2M\omega-k_{z}^{2}}\,.
\end{equation}
\noindent The motion equation (\ref{ed2o}) describes the quantum dynamics of a DKP oscillator in a cosmic string background. With $\phi_{1}(0)=0$ and $\int_{0}^{\infty}dr|\phi_{1}|^2<\infty$, the solution for (\ref{ed2o}) with $\kappa$ and $\lambda$ real is precisely the well-known solution of the Schr\"{o}dinger equation for the harmonic oscillator. The solution close to the origin valid for all values de $m_{\alpha}$ can be written as being proportional to $r^{|m_{\alpha}|+\frac{1}{2}}$. On the other hand, for large $r$ the square-integrable solution behaves as
$e^{-\lambda r^2/2}$, thereby the solution for all $r$ can be expressed as
\begin{equation}\label{ansatz2}
\phi_{1}(r)=r^{|m_{\alpha}|+\frac{1}{2}}e^{-\lambda r^2/2}f(r)\,,
\end{equation}
\noindent subsequently, by introducing the following new variable and parameters:
\begin{eqnarray}
 \rho &=& \lambda r^2\,, \label{newvar}\\
 a &=& \frac{1}{2}\left(|m_{\alpha}|+1-\frac{\kappa^2}{2\lambda}\right)\,,\label{a}\\
 b &=& |m_{\alpha}|+1\,,\label{b}
\end{eqnarray}
\noindent one finds that $f(\rho)$ can be expressed as a regular solution of the confluent hypergeometric equation (Kummer's function) \cite{ABRAMOWITZ1965},
\begin{equation}\label{hipergeome}
\rho\frac{d^2f}{d\rho^2}+\left(b-\rho\right)\frac{df}{d\rho}-af=0\,.
\end{equation}
\noindent The general solution of (\ref{hipergeome}) is given by \cite{ABRAMOWITZ1965}
\begin{equation}\label{gs}
f(\rho)=AM\left(a,b,\rho\right)+B\rho^{1-b}M\left(a-b+1,2-b,\rho\right)
\end{equation}
\noindent where $A$ and $B$ are arbitrary constants. The second term in (\ref{gs}) has a singular point at $\rho=0$, so that we set $B=0$. The asymptotic behavior of Kummer's function is dictated by
\begin{equation}\label{asymp}
M\left(a,b,\rho\right)\simeq \frac{\Gamma(b)}{\Gamma(b-a)}e^{-i\pi a}\rho^{-a}+
\frac{\Gamma(b)}{\Gamma(a)}e^{\rho}\rho^{a-b}\,.
\end{equation}
\noindent It is true that the presence of $e^{\rho}$ in the asymptotic behavior of $M(a,b,\rho)$ perverts the normalizability of $\phi_{1}(\rho)$. Nevertheless, this unfavorable behavior can be remedied by demanding $a=-n$, where $n$ is a nonnegative integer and $b\neq-\tilde{n}$, where $\tilde{n}$ is also a nonnegative integer. In fact, $M(-n,b,\rho)$ with $b>0$ is proportional to the generalized Laguerre polynomial $L_{n}^{(b-1)}(\rho)$, a polynomial of degree $n$ with $n$ distinct positive zeros in the range $[0,\infty)$. Therefore, the solution for all $r$ can be written as
\begin{equation}\label{solg}
\phi_{1}(r)=N_{n}r^{|m_{\alpha}|+\frac{1}{2}}e^{-\lambda r^2/2}L^{|m_{\alpha}|}_{n}(\lambda r^{2})\,,
\end{equation}
\noindent where $N_{n}$ is a normalization constant. The charge density $J^{0}$ (\ref{norma2}) dictates that
$\phi_{1}$ must be normalized as
\begin{equation}\label{norma3}
\frac{|E|}{M}\int_{0}^{\infty}dr|\phi_{1}|^{2}=1\,,
\end{equation}
\noindent so that, the normalization constant can be written as
\begin{equation}\label{norma4}
N_{n}=\sqrt{\frac{2M\lambda^{|m_{\alpha}|+1}\Gamma\left(n+1\right)}{|E|\Gamma\left(|m_{\alpha}|+n+1\right)}}\,,
\end{equation}
\noindent with $|E|\neq 0$. Moreover, the requirement $a=-n$ (quantization condition) implies into
\begin{equation}\label{energyr}
E=\pm\sqrt{M^2+k_{z}^{2}+2M\omega\left(2n+\frac{|m|}{\alpha}\right)}
\end{equation}
\noindent This last result shows that the discrete set of DKP energies are symmetrical about $E=0$ and it is irrespective to the sign of $m$. This fact is associated to that DKP oscillator embedded in a cosmic string background does not distinguish particles from antiparticles. At this stage, due to invariance under rotation along the z-direction, without loss of generality we can fix $k_{z}=0$. In general, $|E|>M$ excepts for $\omega=0$ that the spectrum acquiesces $|E|=M$.

Now, let us consider the weak-coupling limit, $\omega\ll 1$ and $|E|\simeq M$ for small quantum numbers.  With all, the Eq. (\ref{energyr}) becomes
\begin{equation}\label{energy_wcl}
|E|\simeq M\left[1+\omega\left(2n+\frac{|m|}{\alpha}\right)\right]\,,
\end{equation}
\noindent note that due to equally spaced energy spectrum we can say that it describes a genuine DKP oscillator. But, as the weak-coupling limit does not correspond to the nonrelativistic limit, we also can consider the nonrelativistic limit of (\ref{energyr}). Following the standard procedure, $E=M+\mathcal{E}$ with $M\gg\mathcal{E}$, and after some calculations one has that
\begin{equation}\label{lnr}
\mathcal{E}\simeq \omega\left(2n+\frac{|m|}{\alpha}\right)\,.
\end{equation}
\noindent which describes the energy of a traditional nonrelativistic harmonic oscillator.

\begin{figure}[h]
\includegraphics[width=0.93\linewidth,angle=0]{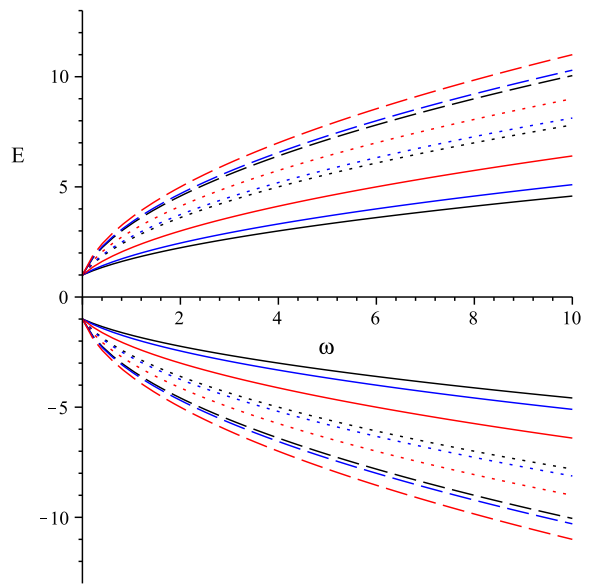}
\caption{\label{fig1} Plots of the energy as a function of $\omega$ for $|m|=1$ and different values of $n$ and
$\alpha$. For $n=0$ [solid line], $n=1$ [dotted line] and $n=3$ [dashed line]. For $\alpha=1$ [black],
$\alpha=0.8$ [blue] and $\alpha=0.5$ [red].}
\end{figure}

\begin{figure}[h]
\includegraphics[width=0.93\linewidth,angle=0]{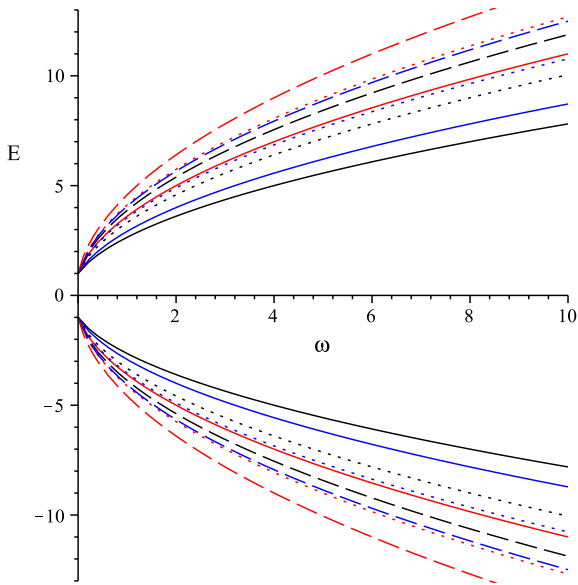}
\caption{\label{fig2} The same as figure \ref{fig1}, for $|m|=3$.}
\end{figure}

Figures \ref{fig1} and \ref{fig2} illustrate the profiles of the energy as a function of $\omega$ for $|m|=1$ and $|m|=3$, respectively. In both figures we consider the three first quantum numbers and three different values for $\alpha$. From figures \ref{fig1} and \ref{fig2} one sees that all the energy levels emerge from the positive (negative)-energy continuum so that it is plausible to identify them with particle (antiparticle) levels. Furthermore, it is noticeable from both of these figures that for positive-energy spectrum one finds that the lowest quantum numbers correspond to the lowest eigenenergies, as it should be for particle energy levels. On the other hand, for negative-energy spectrum presents a similar behavior but the highest energy levels are labeled by the lowest quantum numbers and are to be identified with antiparticle levels.  Also, one can see that for fixed values of $n$ and $|m|$, the energy $|E|$ increases as $\alpha$ decreases.


\begin{figure}[h]
\includegraphics[width=0.93\linewidth,angle=0]{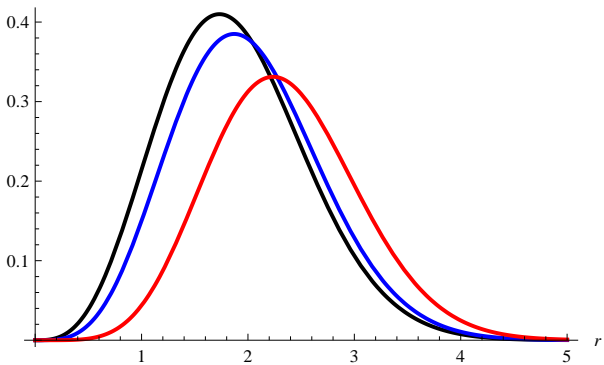}
\caption{\label{fig3} Plots of $|\phi_{1}|^{2}$ for $n=0$, $|m|=1$ and different values of $\alpha$: for
$\alpha=1$ [black], $\alpha=0.8$ [blue] and $\alpha=0.5$ [red].}
\end{figure}

\begin{figure}[h]
\includegraphics[width=0.93\linewidth,angle=0]{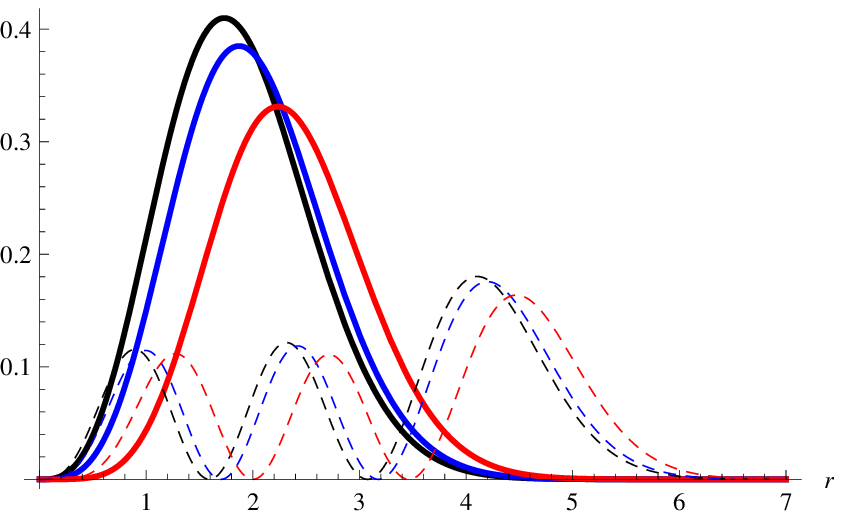}
\caption{\label{fig3_2} Plots of $|\phi_{1}|^{2}$ for $n=0$ [thick line] and $n=2$ [dashed line], $|m|=1$ and different values of $\alpha$: for $\alpha=1$ [black], $\alpha=0.8$ [blue] and $\alpha=0.5$ [red].}
\end{figure}

\begin{figure}[h]
\includegraphics[width=0.93\linewidth,angle=0]{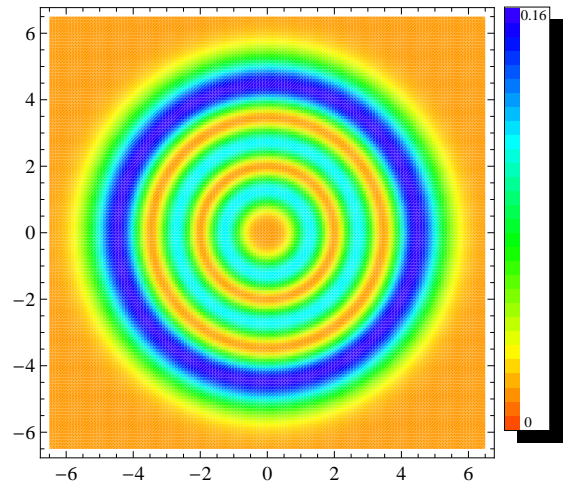}
\caption{\label{fig4} Plots of $|\phi_{1}|^{2}$ for $n=2$, $|m|=1$ and $\alpha=0.5$.}
\end{figure}

In Fig. \ref{fig3}, we illustrate the results of $|\phi_{1}|^{2}$ for $n=0$, $|m|=1$ and different values of $\alpha$. From figure \ref{fig3} one can see that for fixed values of $n$ and $|m|$, the distribution has a maximum at $r\approx 1.7$ for $\alpha=1$, this maximum decreases and moves to positive r-direction as $\alpha$ increases. In addition, comparison between $|\phi_{1}|^{2}$ shows that $\alpha=1$ tends to be better localized than $\alpha<1$. From this, we can conclude that in the limit $\alpha\rightarrow 0$ one has $N_{n}\rightarrow 0$, so that the solution $\phi_{1}$ tend to disappear one after ano\-ther as $\alpha\rightarrow 0$. The comparison between the profiles of $|\phi_{1}|^{2}$ for $n=0$ and $n=2$ are shows in Fig. \ref{fig3_2} for $|m|=1$ and different values of $\alpha$. The Fig. \ref{fig3_2} clearly shows the effects of $\alpha$ on the excited modes, which are qualitatively similar to $n=0$. Finally, Fig. \ref{fig4} illustrates the behavior of $|\phi_{1}|^{2}$ for $n=2$, $|m|=1$ and $\alpha=0.5$ in polar coordinates. One can see that scalar bosons tend to be better localized at the blue region.


\section{Conclusions}
\label{sec:4}

We studied the Duffin-Kemmer-Petiau (DKP) equation in a curved space-time and we found the general condition on the interactions which leads to a conserved current. This result is a generalization of \cite{JPA43:055306:2010} (Minkowski space-time). Furthermore, we showed that considering a cosmic string background and a DKP oscillator interaction, they furnish a conserved current.

Considering only scalar bosons, we showed that the motion equation which describes the quantum dynamics of a DKP oscillator in a cosmic string background was mapped into a Schr\"{o}\-dinger-like equation embedded in a three-dimensional harmonic oscillator for the first component of the DKP spinor and the remaining components were expressed in terms of the first one in a simple way. Our result is very similar to Dirac oscillator in a cosmic string background, except by the absence of some terms that depend on the spin projection parameter \cite{EPJC74:3187:2014}.

We found the spectrum of energy for this background and we showed that the energy $|E|$ increases as $\alpha$ decreases. Both particle and antiparticle energy levels are members of the spectrum, and the particle and antiparticle spectra are symmetrical about $E=0$. That fact implies that there is no channel for spontaneous boson-antiboson creation. We also found that both, weak-coupling limit and nonrelativistic limit furnish equally spaced energy spectrum, so that we concluded that this problem describes a genuine DKP oscillator.

The behavior of the solutions for this problem was discussed in detail. We showed that the cosmic string background influences on the scalar bosons localization. As a important result, we showed that $\alpha=1$ tends to be better localized than $\alpha<1$ (see Fig. \ref{fig3}). Also, we showed that in the limit $\alpha\rightarrow 0$ the solution $\phi_{1}$ tend to disappear.

Beyond to investigate the quantum dynamics of scalar bosons in a cosmic string background, the results of this paper could be used, in principle, in condensed matter physics, owing to the analogy between cosmic strings and disclinations in solids \cite{NELSON2002}. Another physical application could be associated to Bose-Einstein condensates (BEC) \cite{PLA316:33:2003,PA419:612:2015} and neutral atoms. It is known that condensates can be exploited for building a coherent source of neutral atoms \cite{PRL78:582:1997}, which in turn can be used to study entanglement and quantum information processing \cite{PRL82:1975:1999}.

Finally, it is worthwhile to mention that the natural extension of the present work is consider more general backgrounds, as for instance a global monopole \cite{CQG23:5249:2006} and a spinning cosmic string \cite{AP201:241:1990,AP350:105:2014}, among others.

\begin{acknowledgements}
The author would like to thank Edilberto O. Silva for fruitful discussions. Thanks also go to the referee for useful comments and suggestions. This work was
supported in part by means of funds provided by CNPq, Brazil, Grants No. 455719/2014-4 (Universal) and No. 304105/2014-7 (PQ).
\end{acknowledgements}

\bibliographystyle{spphys}       
\bibliography{mybibfile_stars2}

\end{document}